# Relational Access Control with Bivalent Permissions in a Social Web/Collaboration Architecture

Todd Davies and Mike D. Mintz

*Symbolic Systems Program, Stanford University, Stanford, CA 94305-2150 USA*
davies@csli.stanford.edu, mikemintz@cs.stanford.edu

**ABSTRACT**

*We describe an access control model that has been implemented in the web content management framework "Deme" (which rhymes with "team"). Access control in Deme is an example of what we call "bivalent relation object access control" (BROAC). This model builds on recent work by Giunchiglia et al. on relation-based access control (RelBAC), as well as other work on relational, flexible, fine-grained, and XML access control models. We describe Deme's architecture and review access control models, motivating our approach. BROAC allows for both positive and negative permissions, which may conflict with each other. We argue for the usefulness of defining access control rules as objects in the target database, and for the necessity of resolving permission conflicts in a social Web/collaboration architecture. After describing how Deme access control works, including the precedence relations between different permission types in Deme, we provide several examples of realistic scenarios in which permission conflicts arise, and show how Deme resolves them. Initial performance tests indicate that permission checking scales linearly in time on a practical Deme website.*

## 1. INTRODUCTION

Cloud computing and social web applications are increasingly being used by formal organizations (with paid staff and/or hierarchical reporting relationships) as well as informal social networks and virtual communities. In between are groups such as neighborhood associations, ad-hoc citizen groups that organize around particular causes, and committees of members in larger organizations such as schools, universities, and places of worship. In this paper, we argue that groups across this full spectrum, when they use shared computing environments, face issues related to what information each person should be permitted to access. In addition, we argue that previously developed models of access control are inadequate for addressing some of the issues that commonly arise in contemporary social web applications, in which many users have the ability to form overlapping groups and to publish and label data aimed at different groups of users.

Our work on access control is instantiated in a content management framework for building social websites: *Deme* (which rhymes with "team"). Deme is free/open source software designed to facilitate collaborative production, document-centered discussion, and group decision making, flexibly across many different types of organizations of the types mentioned above. The access control scheme introduced in this paper, which we call *bivariate relation object access control* or *BROAC*, grew out of practical problems we faced in creating Deme as a framework that could support closed as well as open groups, with a high degree of user control over permissions to access and perform operations on data in a social web environment.

## 2. THE *DEME* SCHEME

In this section, we describe the architecture of Deme,[1] a web content management system and framework written in Django/Python, with a PostgreSQL database, licensed under the Affero GPLv3 license.[2] Recently, the term *content management framework* has been used, somewhat controversially, to denote "an application programming interface for creating a customized content management system".[3] We use the term *framework* to indicate that the system is designed to facilitate custom code development. Deme attempts to make available the concepts of object-oriented programming (OOP) to end users and nonprogrammer website administrators, using terminology that we believe will be more understandable to nonprogrammers. We define the basic vocabulary of Deme below with respect to concepts familiar to a technical audience.

---

1 See [6].
2 See [14].

**Items and item types.** Units of content in Deme are stored in "items". An Item is an instance of a particular "item type". Deme item types are defined in an *inheritance hierarchy*. If the Person item type inherits from the Agent item type, then any Item that is a Person is also an Agent. Every item type ultimately inherits from the Item item type (which corresponds to the Object class in many programming languages). We allow multiple inheritance, and use it occasionally (e.g., TextComment inherits from both Comment and TextDocument). Deme items are stored in a database using object relational mapping (ORM)[4] with multi-table inheritance. For example, if our item type hierarchy is Item -> Agent -> Person, and our items are Mike[Person] and Robot[Agent], then there will be one row in the Person table (for Mike), two rows in the Agent table (for Mike and the robot), and two rows in the Item table (for Mike and the robot). An abridged basic view of the Deme item type hierarchy is shown in Figure 1.

**Pieces.** Every item type defines the "pieces" (mapped to fields/columns in the database) relevant for that type's Items, and item types inherit pieces from their supertypes. If Item defines the description piece, Agent defines no new pieces, and Person defines the first_name piece, then every person has a description and a first_name.

**Piece types.** Every piece of an Item has a type (e.g. String, Integer, or Boolean). Pieces can point to other Items (foreign keys in the database). Pointing pieces are useful for defining relationships between items. For example, the Item item type has a creator piece pointing to the Agent that created it. Multiple Items can point to a common Item. Pieces cannot store data structures such as lists [8]. So rather than storing different contact methods as pieces of each Agent, we make ContactMethod an item type, and give it an agent_pointer piece. The contact methods for agent

---
[3] See for example [25].
[4] See Scott W. Ambler's explanation of ORM [2].

123 are represented by all of the ContactMethods that have agent_pointer equal to 123.

There are a few types of data object in Deme that are not full Items, but are rather what might be called *quasi-items*. These include Versions (one for each version of an item, archived similar to a wiki), ActionNotices (which record information about actions on an item resulting in a revision), and Permissions (see section 5). These Objects are not Items because to make them so would put us in an infinite regress (since all Items have associated Versions, ActionNotices, and Permissions.

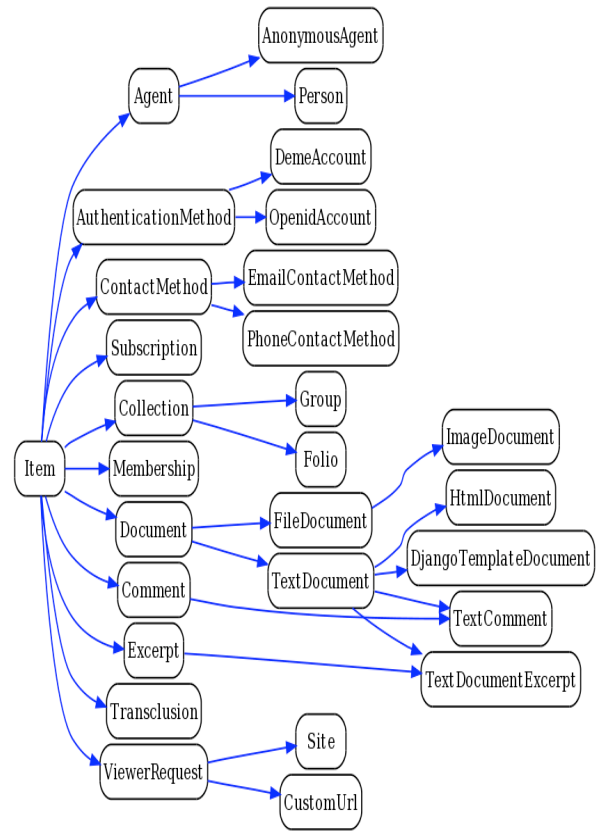

**Figure. 1. The Deme Item Type Hierarchy**

Deme is an architecture for collaboration and the social Web. Every Item is commentable, and Comments can be *transcluded,* along with their resulting threads, at any location in a document, facilitating document-centered discussion [5]. This is a generalization of the idea behind wikis, in that every Item can be set for open editing, but it is also possible to define elaborate (fluid, granular) access control for Items and their associated pieces/fields, in cases such as static websites and social communities where not everything is publicly editable. The framework is very extensible, so that a programmer can define new item types within the Deme hierarchy and take advantage of viewers that have been developed for a new type's supertype [4].

## 3. ACCESS CONTROL MODELS

Access control in a software system is the process by which an authenticated user (the *subject*) gets authorized to perform actions on a data *object*. Subjects' abilities to perform actions on objects are represented in different ways in different models. The access control models summarized below represent different paradigms that can be instantiated with substantial variation in different systems. The models are not mutually exclusive: a given system may embody aspects of more than one model (e.g. the main system described in [18]). Nonetheless, they show how thinking about access control has progressed as software has evolved from traditional file systems to contemporary social platforms.[5]

### 3.1 Discretionary Access Control (DAC)

Traditional time-sharing systems, such as the early multiuser operating system Multics [30] and its successor Unix [29], store data in files, each of which has an owner. These systems give individual subjects the power to create data objects (files in this case) and to grant access to them by other subjects. This model is termed *discretionary access control* or *DAC* because it gives discretion to individual users over who else can access the data they create or "own".[6]

The permissions or rights that subjects have with respect to objects can be represented as an *access control matrix (ACM)*. In an ACM, subjects $s$ are the rows of the matrix, objects $o$ are the columns, and each cell defines the *protection state* of access rights $A[s,o]$ for a subject-object pair [20, 38]. An access control matrix is useful abstractly, but in practice access rights are defined too sparsely and the matrix is too large to be represented explicitly as a table [7]. DAC systems are therefore typically implemented in one of two ways:

- *Capability-based security*, in which a subject obtains access to an object by being given a key or reference which gives that subject the capability to access the data object [23]. This is often represented as a *capability list (C-list)* of subjects who have an access right to for a given object [7, 22].

- *Access control lists (ACLs)*, which represent the set of users/subjects who can access a data object, e.g. those listed as Owner, Group, or Others for a file in the Unix system, together with what types of actions they can perform (e.g. (R)ead, (W)rite, or e(X)ecute) [7, 29].

### 3.2 Mandatory Access Control (MAC)

*Mandatory access control* [39] is a nondiscretionary model in which access cannot be passed from one user to another, but is instead defined by predetermined subject attributes (e.g. security clearances) and object security levels that are enforced by the system. MAC systems came about as a response to the security requirements of the military and other hierarchical/governmental organizations.

### 3.3 Role-Based Access Control (RBAC)

In *role-based* security systems, subjects are given permission to perform actions based on their roles in an organization [9, 31]. RBAC is compatible with either discretionary or mandatory authorization, but in its pure form it represents permissions as features of a role, rather than of an object as an access control list does. A second innovation in RBAC is the ability to represent permissions in a more granular manner, so that a role can define what types of operations can be performed within an object, not just on the object as a whole (e.g. updating one field in a database record versus having write permission on a whole file). This supports a separation of duties in organizations [19, 24]. The RBAC model has been standardized by the National Institute of Standards and Technology (NIST) [10, 32].

The term *role* is often used interchangeably with *group*, as in a "group of users/subjects". Roles or groups can be represented as subjects in an ACM-based implementation. For example, in the *Suite* framework for collaboration [7, 34], permissions are represented in ACLs which can include groups/roles as subjects and groups of objects as objects. Suite was augmented with features for metalevel access control, yielding a powerful tool that can support multiple ownership of objects, rules governing transfer of access rights, negative permissions, and more [7].

---

[5] The access control models described in this section are not exhaustive. A few of the other models proposed include attribute-based access control [40], context-based access control [3, 11], group-based access control [15, 34, 35], task-based access control [1, 37], team-based access control [37], and user-centric access control [28]. In general, these are variants or extensions of the models presented here. See [18] and [38] for a discussion of models related to collaborative systems. For further discussion of the model proposed in [35], see section 4 of this paper.
[6] See [39] for a widely used definition.

### 3.4 Relational Access Control (RAC)

Theoretical work on access control has led to several frameworks for representing access control policies in flexible, fine-grained ways, e.g. [12, 17, 27]. Rather than assuming a particular policy model, such as DAC, MAC, or RBAC, these frameworks typically provide a way to represent *access control rules (ACRs)* [22] that may apply to individual users, roles, and groups, as well as to objects, fields, and collections of objects.

ACRs are relations, each between a subject (which may be a group of subjects), an object (which may be a collection of objects), an action, and sometimes a sign, where the sign (positive or negative) defines whether the subject(s) is (are) permitted or prohibited to perform the action on the object(s). This obviously has expressive efficiency advantages over having to define permissions over singletons, as in a traditional ACL. Granularity may be represented through multiple action types, granular objects, or both.

Different frameworks allow different types of expressions, which may include the ability to infer permissions hierarchically and/or to define metalevel rules. Access control rules may be represented as a set of propositions in a framework or language for specifying ACRs outside of the target database [17, 26]. Alternatively, they may be represented as *relation objects* in the database itself, wherein each row of the permissions table is a rule or relation, rather than a subject as it would be in an ACM-based approach. This overcomes the problem of representing many null values in the usually sparse ACM, and allows more than one ACR to be defined for a given subject-object pair. It also avoids having to store complex structures like lists or arrays within a table cell, in keeping with standard relational database practice [8]. We call this model *relation object access control (ROAC)*.

ROAC's main advantage over outside-the-database methods of relational access control is that it integrates permissions within the database, so that code designed to interact with objects can access permissions/ACRs as well. In a social Web/collaboration context, this is useful because ACRs may need to be displayed in the context of the subjects and objects to which they refer, as well as searched and discussed. Using the database to represent permission relations can also be efficient, because pointers in the database can be used to refer to data objects in the ACR. And perhaps most importantly, it allows users themselves to modify permissions dynamically within the normal user interface.

An example of ROAC was recently proposed by Giunchiglia et al., motivated specifically by new contexts such as grid computing, social web applications, and semantic desktops [12, 13]. In their system, permissions are represented separately from both the subject and the object, in a relation which points to both and which has the status of a first class object. Giunchiglia et al. call their model *relation-based access control* or *RelBAC*.

The core of the RelBAC approach is the representation of a permission as a relation object pointing to a subject and a data object. This is motivated by a need to manipulate and reason about permissions independently from users and the other stored data in a system, which in software with multiple instances will differ from one installation to another. But the RelBAC model itself is more specific than the ROAC model as we have described it. In addition to SUBJECT (or USER), OBJECT, and PERMISSION (an operation the user can perform on the OBJECT), the entity relationship model for RelBAC defines hierarchical sets of USERS (GROUPS) and of data OBJECTS (CLASSES) through an IS-A relation, as well as ACCESS CONTROL RULES and POLICIES, which instantiate permissions to specific groups and classes.

The features of RelBAC are all theoretically useful, as they allow for inference of permissions from other permissions, and of subgroup permissions from supergroup permissions, among other capabilities. However, the more complex features of the model rely for their usefulness on assumptions that may not hold empirically in a given context, for example about the existence of hierarchical relationships among groups and among classes, which may in reality be overlapping rather than hierarchical. RelBAC also does not incorporate negative permissions/prohibitions (see section 4). The additional complexity of run-time reasoning required to make use of RelBAC's features is formidable and has so far apparently resisted a tractable solution [41]. Because Giunchiglia and colleagues have defined RelBAC in a rather specific way, we believe it is useful to classify it within this larger theoretical class of models we call *ROAC*.

*Relational access control (RAC)* has been the subject of considerable study in contexts where fine-grained permissions must routinely be assigned, e.g. for XML documents [21, 22, 26, 27]. But to our knowledge the RAC paradigm has not yet been recognized as a significant alternative for collaboration systems.[7] In what follows, we try to demonstrate how the ROAC variant of RAC can be implemented practically, as well as some of its advantages in a collaborative context.

## 4. BIVALENT PERMISSIONS

Traditional access control systems define access only in positive terms. If a subject is on an access control list or has a capability/key for an object, then the subject can access the object. Otherwise, and by default, the subject cannot access the object. But this does not allow for possible conflicts that can arise in social or collaborative contexts, in which an organization might want to explicitly prohibit some individuals from having access to a given data object when their membership in a group would otherwise give them access. An organization might want, for example, to exclude a member of its board of directors from accessing reference letters that were consulted by the board when it invited that member to join the board. Examples like this abound (see section 6).

The need to define negative permissions (prohibitions) in some form, either explicitly with the need to resolve conflicts (e.g. [34]) or via constraints in a role-based model (e.g. [16]) has been recognized by researchers studying collaborative systems going back at least to the 1970s [30]. As Sikkel [35] notes: *"A straightforward model of negative rights is used in the Andrew system [33]. Rights exist both in positive and negative form. When both apply, the negative right overrides the positive right. In this way, negative rights can be used to immediately revoke a permission in a distributed system where propagation of changes may take a while."* Although some systems such as Suite [7, 34] have utilized more complicated conflict resolution rules, negative permissions typically override positive ones [22].

We call models that allow both permissions and prohibitions *bivalent*. In the access control system defined below, we too resolve conflicts between negative and positive permissions by giving precedence to the negative. Some motivating examples and how they are resolved in our system are given in section 6.

---

[7] See the 2005 overview of access control models for collaborative systems by Tolone et al. [38] and section 2 of the 2006 paper by Kim et al. [18], neither of which mentions relational approaches.

## 5. THE DEME PERMISSIONS SYSTEM

Deme implements an access control model in which permissions are defined as relation objects pointing to a subject and an object. It is therefore an instance of what we have called *relation object access control (ROAC)*. Permissions can also be either positive or negative. To our knowledge, Deme is the first instance of such a *bivalent ROAC (BROAC)* model, and the first implementation of ROAC in a real-world context.

In what follows, we describe the particular access control procedures in Deme. We intend *BROAC* to refer to a larger class – any model in which access control is both bivalent and represented as relation objects in the target database.

Permissions in Deme define what actions Agents can and cannot do. Similar to ActionNotices, permissions are not Items themselves, but they are quasi-items that exist in the database and point to Items.

There are 9 types of permissions in Deme, divided along 2 axes: the subject (data seeker) axis and the object (data sought) axis. Along the subject axis, permissions can be given at 3 levels: to a single Agent, to the members of a Collection of Agents, or to all Agents. Along the object axis, permissions can be applied to 3 levels: to a single Item, to the Items in a Collection, or to all Items. For both axes, we refer to these three levels as "one", "some", and "all". The 9 permission types are shown in Table 1.

**Table 1. The Deme Precedence System - Numbers in Parentheses Refer to Precedence Order**

|  |  | *Item* | *Collection* | *All Items* |
|---|---|---|---|---|
| **Subject** | *Agent* | One To One **(1)** | One To Some **(2)** | One To All **(3)** |
|  | *Group* | Some To One **(4)** | Some To Some **(5)** | Some To All **(6)** |
|  | *All Agents* | All To One **(7)** | All To Some **(8)** | All To All **(9)** |

(Object column headers)

Although we could accomplish anything using only OneToOnePermissions, the other permission types allow us to more concisely express permissions. For example, if our site was a wiki and we wanted any user to be able to edit any document, we would create a single AllToAllPermission, rather than a new OneToOnePermission for every Agent/Item pair.

Each permission, in addition to specifying the subject and the object axes, specifies an ability string and an is_allowed boolean. When there are multiple permissions with the same ability, the permissions at a level with a lower number (shown in parentheses after each permission type in Table 1) take precedence. When there are multiple permissions at the same level, the negative (is_allowed=False) permissions take precedence over the positive permissions. Two permissions referencing the same subject, object, and ability cannot differ only in the value of is_allowed. Access control in Deme embodies a closed world assumption: Access is not allowed unless it is positively permitted by at least one permission object.

The general principles that (a) prohibitions override (positive) permissions and that (b) more explicit access rules override less explicit ones are fairly standard [22], but the exact ordering of the precedence system might be controversial. For now, we regard the precedence system as plausible but in need of empirical validation by users.

On both axes, when we refer to all Agents or Items in a collection (i.e., [X]ToSome or SomeTo[X]), we refer to both direct and indirect members. Thus, Deme checks the RecursiveMembership table to determine whether an Agent or an Item is affected by the permission.

There are two types of abilities: item abilities and global abilities. Item abilities can apply to a particular Item (or Collection of Items), such as "can edit the name of the Item"; but global abilities apply to Items generally, e.g. "can create new Documents". Each item type defines the item abilities that are relevant to it, and the global abilities it introduces. An Agent has an ability if (a) there exists a relevant permission with is_allowed=True at some level and (b) there are no relevant permissions with is_allowed=False at any levels with the same or lower precedence number.

The global abilities defined in Deme are given in Table 2.[8]

---

[8] Agents with the do_anything ability automatically have every single global ability and every item ability with respect to every Item. If an agent has this global ability in the final calculation, this overrides any item abilities at any level. As a specific unusual example, if an agent has the global do_anything ability from an EveryonePermission, then giving him/her any item ability with is_allowed=False will have no effect.

**Table 2. Current Global Abilities in Deme**

create AIMContactMethod
create AddressContactMethod
create Agent
create Collection
create CustomUrl
create DemeAccount
create DjangoTemplateDocument
create EmailContactMethod
create Event
create FaxContactMethod
create FileDocument
create Group
create HtmlAdvertisement
create HtmlDocument
create ImageDocument
create Membership
create Person
create PhoneContactMethod
create Subscription
create TextAdvertisement
create TextComment
create TextDocument
create TextDocumentExcerpt
create Transclusion
do_anything

Some sample item types in Deme and the abilities they introduce are shown in Table 3.

Deme implements discretionary access control: When an Item is created, by default no permissions are created except a OneToOne, do_anything permission between the creator and the Item. With this ability, the Item creator can create whatever other permissions s/he wants, either during the creation process, or later.

When a database query is processed, Deme takes the currently authenticated Agent and decides whether the Agent has the required ability to complete the requested action (or to display some part of the view). Abilities are not just checked before doing actions, but they can also be used to filter out Items on database lookups. For example, if a viewer is supposed to display a list of Items the user is allowed to see (because they have the view Item.name ability), it will need to use permissions to filter out inappropriate results.

To modify an [X]ToOne permission, one must have the do_anything ability with respect to the target Item. Similarly, to modify an [X]ToSome permission, one must have the do_anything ability with respect to the target Collection. Finally, to modify an [X]ToAll permission, one must have the global do_anything ability.

**Table 3. Some Item Abilities in Deme**

| Item type | Item abilities defined by the item type |
|---|---|
| Item | do_anything<br>comment_on<br>delete<br>view Item.name<br>view Item.description<br>view Item.creator<br>view Item.created_at<br>edit Item.name<br>edit Item.description |
| Agent | add_contact_method<br>add_authentication_method<br>login_as<br>view Agent.last_online_at |
| Person | view Person.first_name<br>view Person.middle_names<br>view Person.last_name<br>view Person.suffix<br>edit Person.first_name<br>edit Person.middle_names<br>edit Person.last_name<br>edit Person.suffix |
| Collection | modify_membership<br>add_self<br>remove_self |
| Text Document | view TextDocument.body<br>edit TextDocument.body<br>add_transclusion |
| Site | view Site.hostname<br>edit Site.hostname<br>view Site.default_layout<br>edit Site.default_layout |

There is a potential loophole in the setup described above. A user could create a Collection, add a private Item to it (because they have do_anything with respect to the Collection), create an [X]ToSome permission for that Collection (because they have do_anything with respect to it), and thus gain full access to the private Item. In order to resolve this, we use the permission_enabled field in Membership. [X]ToSome permissions only propagate to members of the Collection through Memberships with permission_enabled=True, and Agents can only modify the permission_enabled field of a Membership if they can do_anything to the member Item.

By enforcing this, we guarantee that when a user modifies an [X]ToSome permission, it only affects Items in the Collection that were added to it with permission_enabled=True, by a user with power over that Item. Since [X]ToSome permissions recursively traverse Memberships, we have a permission_enabled field in RecursiveMembership that is set to True if and only if there exists a path of Memberships from the parent Collection to the child Item, all with permission_enabled=True.

Deme allows (optionally) for anyone to view content through the agent Anonymous. This creates another potential loophole for users who have negative permissions with respect to an Item, i.e. the user Anonymous may have a positive permission with respect to the Item, so that the user could simply log out and do what they are not supposed to be able to do. The interface must convey this to users who create negative permissions, so that anonymous access is disallowed in a similar fashion.

## 6. EXAMPLES OF PERMISSION CONFLICT RESOLUTION

To appreciate the variety of conflicts that can arise in a system like Deme, and the way in which the version of BROAC implemented in Deme resolves them, we offer the following example scenarios.

**Example 1.** The executive director of a nongovernmental organization, who is hired and supervised by the NGO's board of directors, has access to most board documents as a member of the board's Group, but does not have access to those documents related to the board's deliberations over the executive director himself. The board's Group permission for reading its Folio is positive for the Collection of executive director hiring and review documents. The executive director's Agent permission for reading this Collection is negative. The latter (negative) permission has precedence. *2(-) defeats 5(+)*.

**Example 2.** Each student has access to their own transcript, but not to those of other students. The Group of students has a negative permission for reading a student's transcript. But a student's Agent permission is positive for reading their own transcript. The latter (positive) permission has precedence. *1(+) defeats 4(-)*.

**Example 3.** A student is a programmer for an academic program, and also a member of the staff Group as well as the Group of students. The staff Group has a positive permission for reading student intern applications. The students Group has a negative permission for reading intern applications. The latter (negative) permission has precedence, reflecting a policy that students cannot view transcripts of other students, regardless of their staff status. *5(-) defeats 5(+)*.

**Example 4.** The personnel manager at a firm is a member of the staff Group. The staff has a negative Group permission for accessing staff salary documents. But the personnel manager has a positive Agent permission for accessing salary documents. The latter (positive) permission has precedence. Note, however, that this example shows a limitation in Deme: that we cannot have group-based precedence relations. *2(+) defeats 5(-)*.

**Example 5.** A member of a grassroots advocacy group edits the organization's homepage, publishing a statement not approved by the group. The member has been given permission to edit the homepage, but the group asks the webmaster to place a hold on the offending member's editing ability until the group can discuss the situation. The member has a prior positive Agent permission to edit the homepage. A new, negative Agent permission will replace, rather than coexist with, the old permission. No conflict is possible in this case.

**Example 6.** The webmaster for an organization has the do_anything ability, and can therefore read any document, but is also a member of the staff. The Group for the staff has a negative permission on a security codes document. But the webmaster's Agent permission is positive for all Items, and therefore for this document. The latter (positive) permission has precedence. *3(+) defeats 4(-)*.

**Example 7.** The user of a social networking site hosted on Deme makes a photograph prohibited to all users by default, but grants permission to see her photos to her Group of friends. All users have a negative permission on the forbidden photograph. But the Group of friends has a positive permission for the user's photographs, including the forbidden one. The latter (positive) permission has precedence. This illustrates how users both should be aware of and can take advantage of the precedence relations. *5(+) defeats 7(-)*.

**Example 8.** A user of a Deme social networking site sees several photographs of herself in another user's Collection, the Items in which are visible to all users. She decides that she does not want those photos to be seen by anyone except herself. So she adds all the photos she does not want seen to her own Collection, labeled "private", and sets a negative permission on the Collection for all users (except herself). The photos are permission_enabled. The latter (new, negative) permission has precedence. *8(-) defeats 8(+)*.

## 7. PERFORMANCE

The system described here (Deme) is still in pre-release, but is already powering four websites, including the beta version of our academic program's website [36].

In order to empirically evaluate the performance of Deme's permissions system, we analyzed our program's beta-version social content management/networking website [36], which contains approximately 1,000 users, 12,000 items, and 24,000 permission objects. We made copies of the site with subsets of the content (one with 10 users and their items, one with 20 users, and so forth). For each copy of the site, we performed a query that retrieved all items whose name is visible to the anonymous user. We ran the query 10 times and calculated the average time for the query, and we also ran an alternative version of the query that just retrieved all items without any permission checking. The results of this analysis can be seen in Figure 2.

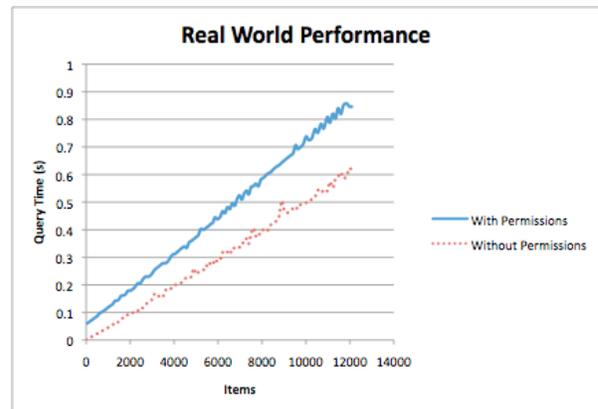

**Figure 2. Performance of Deme Item Access**

Doing a linear regression, we see that in both queries the relationships between website size and query time are linear with R-squared greater than 99%. In the query without permission checking, there is a marginal time of 51 microseconds per item; while in the query with permission checking, the marginal time is 67 microseconds. Of course, this does not necessarily apply in all cases. It depends on the structure of the particular website. In our particular examples, there were about 12 items and 24 permissions for every user.

## 8. CONCLUSION

Deme and its associated permissions system provide a practical example for social/collaboration systems of the application of an emerging paradigm: relational access control. The need for flexible, fine-grained control that is visible to users as a seamless part of the system seems important for this type of platform, as it more easily allows users to interact with and modify permissions within the application itself. We have argued for and implemented a variant of this model – *biavalent relation object access control (BROAC)*, describing how it functions, motivating it through examples, and reporting initial performance tests.

The BROAC approach used in Deme represents a number of compromises. It does not allow specifying that one group or one collection has priority over another for resolving permission conflicts, and it does not allow for specifying hierarchical inference relationships between groups of users or collections of objects. But it does address the broad flexibility of ways that users can define groups and collections which overlap each other, and it accommodates negative permissions, in a relational access control model that we believe to be the first of its kind to be used in a practical system.